\documentclass[traditabstract]{aa}
\usepackage{graphicx}                    
\usepackage{bm}
\usepackage{color}                       
\usepackage{url}                         

\renewcommand{\vec}[1]{\mbox{\boldmath$#1$}}

\begin{document}



\title{The cross helicity  at the solar surface by simulations and observations}
\titlerunning{Cross helicity}
%
\author{G.~R\"udiger\inst{1} \and M.\ K\"uker\inst{1} \and { R.S.}\ Schnerr\inst{2}}


%
  \institute{Leibniz-Institut f\"ur Astrophysik Potsdam, An der Sternwarte 16, D-14482 Potsdam, Germany,
                     email:  gruediger@aip.de, mkueker@aip.de
		     \and Dept. of Astronomy, { Stockholm University}, Alba Nova University Center, 10691 Stockholm, Sweden}

\date{Received; accepted}
 
\abstract{The quasilinear mean-field theory for  driven MHD turbulence  leads to the result that the observed 
 cross helicity $\langle \vec{u}\cdot \vec{b}\rangle$  may directly yield  the magnetic eddy diffusivity $\eta_{\rm T}$ of the quiet Sun. 
In order to model  the cross helicity at the solar surface, magnetoconvection under the presence of a vertical large-scale magnetic field is simulated with the nonlinear MHD code NIRVANA. The very robust result of the calculations is  that  $\langle u_z b_z\rangle\simeq 2 \langle \vec{u}\cdot \vec{b}\rangle$ independent of the applied magnetic field amplitude. The correlation coefficient for the cross helicity is about 10\%. Of similar robustness is the finding that the rms value of the magnetic perturbations exceeds the mean-field amplitude (only) by a factor of five.
The characteristic helicity speed $u_\eta$ as the ratio of the eddy diffusivity  and the density scale  height for an isothermal  sound velocity of 6.6 km/s proves to be 1 km/s for weak fields. This value well coincides  with empirical results  obtained from the data of the HINODE satellite and the Swedish 1-m Solar Telescope (SST) providing the cross helicity component $\langle u_z b_z\rangle$. Both simulations and observations thus lead to a numerical value of  $\eta_{\rm T}\simeq 10^{12}$ cm$^2$/s as characteristic for the surface of the {{\em quiet}} Sun.
}


%

\maketitle
%
\section{Introduction}
{ It is not easy to measure the turbulent magnetic diffusivity $\eta_{\rm T}$ at the solar surface}. This quantity determines the decay of magnetic magnetic structures with scales larger than those of the turbulence. Theoretically, the decay of the magnetic structures should depend on the relation of the magnetic field amplitude to the so-called equipartition value $B_{\rm eq}= \sqrt{\mu_0\rho \langle u^2\rangle}$ defined by the turbulence. This phenomenon is known as the effect of $\eta$-quenching, i.e. the suppression of the eddy diffusivity by the magnetic field.

The simplest realization  of the $\eta$-quenching at the solar surface can be given with two numbers. The decay of active regions  after Schrijver \& Martin (1990) can be understood with an eddy diffusivity of $10^{12}$ cm$^2$/s while the decay of sunspots which their much stronger fields leads to $10^{11}$ cm$^2$/s (Stix 1989).  These values are smaller than the value of $3\cdot
10^{12}$ cm$^2$/s which results from the widely used formula $\eta_{\rm
T}\sim c_\eta u_{\rm rms} \ell_{\rm corr}$ with the tuning parameter $c_\eta\simeq 0.3$, the correlation 
length $\ell_{\rm corr}$ and parameter values taken
close to the surface. { Up to now there was no possibility to observe the
turbulent diffusivity} on the solar surface for the quiet Sun where the
magnetic quenching of this quantity by large-scale magnetic fields is
negligible.

R\"udiger et al.\ (2011) have shown that the combination of a vertical field with
a driven turbulence in a density stratified medium leads to an anticorrelation of the  cross helicity and the vertical large-scale field, i.e. $\langle\vec{u}\cdot\vec{b}\rangle=- \eta_{\rm T} B_z/H_\rho$ with $H_\rho$ as the scale height of the density. If both  the cross helicity and the large-scale vertical field are known then the ratio of the eddy diffusivity and the density scale can be computed. If also the density scale is known from calculated atmosphere models then fluctuation measurements   can be used to calculate the numerical value of the eddy diffusivity for weak fields. This the more as     $\langle\vec{u}\cdot\vec{b}\rangle\simeq \langle u_z b_z\rangle$ if the large-scale magnetic field has only a vertical component and the only vertical gradient is due to the density stratification. The correlation of the vertical components of flow and field can empirically be obtained by both Doppler measurements and {spectropolarimetry}.

To estimate  the value of the cross helicity we assume a density scale height of 100 km and write the result in the form
\begin{equation}
\frac{\langle\vec{u}\cdot\vec{b}\rangle}{ B_z }\simeq - \frac{\eta_{12}}{H_7}   \ \ \ \ {\rm km/s},
\label{pred}
\end{equation}
where $H_7= H_\rho/100\ {\rm km}$ and $\eta_{\rm T}= 10^{12}\ \eta_{12}$ cm$^2$/s. With observations of the LHS of (\ref{pred}) of about 1 km/s one would find $\eta_{\rm T}$ of order $10^{12}$ cm$^2$/s. In the present paper numerical simulations of stratified magnetoconvection and  observational results are discussed and the theory will also  be extended to the inclusion of a vertical stratification of the turbulence intensity. Both the simulations as well as the observations lead to very similar results for the desired magnetic eddy diffusivity for the 
{quiet} Sun exceeding the value given by (\ref{pred}) by a factor of (only) two.

A simple prediction of this theory is that that the ratio (\ref{pred}) does not depend on the sign of the mean magnetic field, i.e. it does not vary from cycle to cycle and (for a dipolar field) from hemisphere to hemisphere. 
As a consequence, the sign of the cross helicity $\langle\vec{u}\cdot\vec{b}\rangle$   should vary from cycle to cycle and between the hemispheres. Zhao et al.~(2011) indeed found indications for a variation from hemisphere to hemisphere in  SOHO/MDI magnetograms and dopplergrams recorded in  2000, 2004 and 2007.

 \section{Mean-field electrodynamics}\label{sect1} 
Let $\vec{U} +  \vec{u}$ and
$ \vec{B} + \vec{b}$ be the fluctuating
 velocity and magnetic field with the  average values $\vec{U}$ and $ \vec{B}$.
The scalar correlation
between the fluctuations of flow and field, i.e. the cross helicity
$\langle \vec{u}\cdot \vec{b}\rangle$, is a pseudoscalar. In the same sense, the cross correlation tensor $\langle u_i b_j\rangle$ is a pseudotensor. We are here only interested  on its symmetric part 
\begin{equation}
H_{ij} =\frac{ \langle u_i b_j\rangle +
\langle u_j b_i\rangle}{2}.
\label{H}
\end{equation}
  As we have shown the tensor  $H_{ij}$ can be finite  in presence
of a mean magnetic field $\vec B$ and for density-stratified  fluids (R\"udiger et al. 2011).
Consider these quantities as small enough that expressions linear in the mean magnetic field  influence of these quantities are sufficient. The same may hold for the shear which influences the (radial) magnetic field components.
It is then straightforward to formulate  the relation
\begin{eqnarray}
\lefteqn{H_{ij}
= \alpha ( \vec{G}\cdot \vec{B}) \delta_{ij} + \beta (G_iB_j+G_jB_i) + \gamma (B_{i,j}+ B_{j,i} )+}\nonumber\\  
&&+a ( \vec{g}\cdot \vec{B} )(U_{i,j}+U_{j,i}) +\nonumber\\
&& +b (U_{i,l}B_j+U_{j,l}B_i) g_l
 +c(U_{i,l}g_j+U_{j,l}g_i)B_l +\nonumber\\
&& + d (U_{l,i}B_j+U_{l,j}B_i) g_l
+e (U_{l,i}g_j+U_{l,j}g_i)B_l.
\label{eq2}
\end{eqnarray}
No other formations are possible linear in the mean field $\vec{B}$, the stratification vector  $\vec{G}$ and the shear of the divergence-free mean flow $\vec{U}$. 
For the tensor components we find
\begin{equation}
H_{yz}=\beta g B_y + (a+b+c) g B_z U_{y,z}
\label{yz}
\end{equation}
and
\begin{equation}
H_{zz}=(\alpha+2 \beta) g B_z + 2 e g B_y U_{y,z}
\label{zz}
\end{equation}
if a box coordinate system $(x,y,z)$ for the latitudinal, azimuthal and vertical direction is introduced. The $z$-axis is aligned with the stratification vector, i.e.~represents the radial direction in spherical geometry. The $x$ and $y$ coordinates denote the horizontal directions. Without shear the correlation $H_{zz}$ measures the vertical magnetic field while the correlation $H_{yz}$ measures the azimuthal field. The correlations are also influenced by the  shear $U_{y,z}$. With the shear included   finite values for both the correlations { (\ref{yz}) and (\ref{zz})} result even for the case that the field has only one component. For known values of the correlations, the coefficients and the vertical field both the azimuthal field {\em and} the shear can be computed. We cannot, however, be sure that all the coefficients $a....e$ must be nonzero. First test calculations of $H_{zz}$ under the presencee of horizontal field {\em and} shear did not yield finite values of $e$ (A. Brandenburg, private communication).

The turbulent flow is assumed anelastic, so that $\mathrm{div}\,\rho\vec{u} = 0$.
It is convenient to use the Fourier transformation of the momentum density
$\vec{m} = \rho\vec{u}$, i.e.
\begin{equation}
	\vec{m}(\vec{r},t) = \int\hat{\vec{m}}(\vec{k},\omega)\
        \mathrm{e}^{\mathrm{i}(\bm{k}\cdot\bm{r} 
	- \omega t)}\mathrm{d}\vec{k}\ \mathrm{d}\omega , 
	\label{1}
\end{equation}
and similarly for the fluctuation of the magnetic field.

The spectral tensor of the momentum density that accounts for the stratification of the turbulence  to the first order terms reads
\begin{eqnarray}
	\lefteqn{\langle\hat{m}_i (\vec{z},\omega )\hat{m}_j(\vec{z}',\omega ')\rangle = 
	\delta (\omega + \omega ') \frac{\hat{q}(k,\omega ,\vec{\kappa})}{16\pi k^2}} 
	\nonumber \\
	&&\quad\quad\quad\quad\quad \times\left(\delta_{ij} -k_ik_j/k^2 
	+ \left(\kappa_i k_j - \kappa_j k_i\right) /(2k^2)\right),
	\label{3}
\end{eqnarray}
where $\vec{k} = (\vec{z}-\vec{z}')/2,\ \vec\kappa = \vec{z} + \vec{z}'$, $\hat{q}$ is the Fourier transform of the local spectrum,
\begin{equation}
	q(k,\omega,{\vec r}) = \rho^2E(k,\omega ,\vec{r}) = 
        \int \hat{q}(k,\omega , \vec{\kappa})
	\mathrm{e}^{ \mathrm{i}\bm{\kappa}\cdot\bm{r}}\ \mathrm{d}\vec{\kappa},
	\label{4}
\end{equation}
so that
\begin{equation}
	\langle u^2\rangle = 
	\int\limits_0^\infty\int\limits_0^\infty E(k,\omega ,\vec{r})\ 
	\mathrm{d}k\mathrm{d}\omega .
	\label{5}
\end{equation}
Derivation of the cross correlation yields
\begin{eqnarray}
\lefteqn{  H_{ij} = \frac{1}{2}\eta_\mathrm{T}
	\left(G_i  B_j + G_j  B_i \right)}\nonumber\\ 
		&&\quad\quad\quad\quad -\left( \frac{3}{10}\eta_\mathrm{T} + \frac{2}{15}\hat{\eta}\right)
	\left( B_{j,i} +  B_{i,j} \right) ,
	\label{6}
\end{eqnarray}
where $\vec{G} = \vec{\nabla}\mathrm{log}\rho $  is the gradient of density  and
\begin{eqnarray}
   \eta_\mathrm{T} &=& \frac{1}{3}\int\limits_0^\infty\int\limits_0^\infty
	\frac{\eta k^2 E}{\omega^2+\eta^2k^4}\mathrm{d}k\ \mathrm{d}\omega ,\label{7.1}\\
	\hat\eta &=& \int\limits_0^\infty\int\limits_0^\infty 
	\frac{\eta k^2\omega^2 E}{(\omega^2+\eta^2k^4)^2}
	\mathrm{d}k\ \mathrm{d}\omega ,
\end{eqnarray}
where $\eta$ is the molecular magnetic diffusivity.
Both quantities remain finite in the  high-conductivity limit. 

From the cross correlation tensor (\ref{6})  the cross helicity 
$
	\langle\vec{u}\cdot \vec{b}\rangle = 
	\eta_\mathrm{T}\left(\vec{G}\cdot  \vec{B}\ \right) $
is obtained.
From Eq.~(\ref{6}) we find the slightly more complicated expression
\begin{equation}
	\langle u_z b_z\rangle = \eta_\mathrm{T} G  B_z 
		-\left(\frac{3\eta_\mathrm{T}}{10} + \frac{2\hat{\eta}}{15}\right)
	\left(2\frac{\partial  B_z}{\partial z}\right) ,
	\label{9}
\end{equation}
where $G=G_z$ is the only nonzero radial components of the density-stratification vectors.
Note the negativity of $G$.
An upward divergence of the mean field would reduce the effect of density
stratification but for uniform field components the result is $\langle u_z b_z\rangle=\langle\vec{u}\cdot \vec{b}\rangle$. 

A real difference, however, between the both  correlation expressions is due to a possible gradient $\vec{G}'$ of the turbulence intensity $u_{\rm rms}$. One easily finds 
that for  vertical fields the turbulence intensity gradient $\vec{G}'$ enters the expressions for the correlations such as
\begin{equation}
 \langle\vec{u}\cdot \vec{b}\rangle = (G+\frac{1}{2} G')\eta_{\rm T} B_z,\ \ \ \ \  
 \langle u_z b_z\rangle =(G+\frac{3}{10} G')\eta_{\rm T} B_z. 
\label{10}
\end{equation}
In the bulk of the convection zone  $G'=G'_z$ is positive while    $G$ is negative. Hence, $|\langle u_z b_z\rangle|>|\langle\vec{u}\cdot \vec{b}\rangle|$ for positive $B_z$ which is confirmed by the presented simulations (see below).

By elimination of $G'$ one finds
\begin{equation}
\frac{5}{2}\langle u_z b_z\rangle-\frac{3}{2}  \langle\vec{u}\cdot \vec{b}\rangle = -\frac{\eta_{\rm T}}{H_\rho} B_z. 
\label{12}
\end{equation}
The magnetic eddy diffusivity can thus be determined if the LHS of
(\ref{12}) is calculated from magnetoconvection simulations  when the density scale height $H_\rho$ is known
from numerical models of the solar atmosphere. As only the correlation $\langle u_z b_z\rangle$ can directly be observed  one needs a numerical model for the application of the LHS of (\ref{12}) to derive the eddy diffusivity at the solar surface. 
\section{Numerical simulations}
We perform simulations for a number of different parameter combinations.
These parameters  include the strength of the imposed
vertical field $B_z$, the viscosity  $\nu$ and  the magnetic diffusivity coefficient $\eta$.

{
The numerical simulations are done using the Nirvana code, which uses a conservative finite difference scheme (Ziegler 2004). We use Cartesian coordinates. 
The code solves the equation of motion, 
\begin{eqnarray} 
  \frac{\partial (\rho \vec{u})}{\partial t} + \nabla \cdot \left[ \rho \vec{u} \vec{u}
  + \left(p + \frac{1}{8 \pi}|\vec{B}|^2\right) I - \frac{1}{4 \pi} \vec{B} \vec{B} \right] = \nonumber \\
   \nabla \cdot \tau + \rho \vec{f}_e, \label{motion}
\end{eqnarray}
the induction equation, 
\begin{equation}
  \frac{\partial \vec{B}}{\partial t} - \nabla \times (\vec{u}\times\vec{B} - \eta \nabla \times \vec{B} ) = 0,
\end{equation}
the equation of mass conservation,
\begin{equation}
  \frac{\partial \rho}{\partial t} + \nabla \cdot (\rho \vec{u}) = 0,
\end{equation}
and the equation of energy conservation,
\begin{eqnarray}
   \frac{\partial e}{\partial t} + \nabla \cdot \left[\left(e+p+\frac{1}{8 \pi} |\vec{B}|^2\right)\vec{v} - \frac{1}{4\pi} (\vec{u}\cdot \vec{B}) \vec{B} \right]  = \nonumber \\
    \nabla \cdot \left[ \vec{u} \tau + \frac{\eta}{4\pi} \vec{B} \times(\nabla \times  \vec{B}) - \vec{F}_{\rm cond} \right] +\rho  \vec{f}_{\rm e} \cdot \vec{u}.  \label{energy}
\end{eqnarray}
In Eqs.~\ref{motion} and \ref{energy}, $\vec{f}_e$ the (external) gravity force and 
\begin{equation}
  \tau=\nu (\nabla \vec{u} + (\nabla \vec{u})^{\rm T} - \frac{2}{3} (\nabla \cdot \vec{u}) I) 
\end{equation}
the viscous stress tensor. The total energy density is the sum of the thermal, kinetic, and magnetic energy density:
\begin{equation}
  e=\epsilon + \frac{\rho}{2} \vec{u}^2 + \frac{1}{2 \mu} \vec{B}^2.
\end{equation}
We assume an ideal gas with a constant mean molecular weight $\mu=1$. The thermal energy density is then
\begin{equation}
  \epsilon = \rho T \frac{\cal R}{\gamma-1}.
\end{equation}   
with $\gamma=c_p/c_v=5/3$.
The gas is heated from below and kept at a fixed temperature at the top of the simulation box. Periodic boundary conditions apply at the horizontal boundaries. A homogeneous vertical magnetic field is applied. The upper and lower boundaries are impenetrable and stress-free. 
}
  
The simulation volume is a rectangular box. The stratification is along the $z$- coordinate and it is piecewise polytrophic, with the polytrophic index chosen such that the hydrostatic equilibrium state is convectively stable in the lower and unstable in the upper half of the simulation box. 
In the following, $p$ denotes gas pressure, $\rho$ mass density, $T$ temperature, $g$ gravity, $\kappa$ thermal conductivity and $c_p$ the specific heat capacity at constant pressure.
 
The gas is initially in hydrostatic equilibrium, i.e.
\begin{equation}
\frac{\partial p}{\partial z} + \rho g =0,
\end{equation}
where $g={\rm const.}$,  and the heat flux through the box is vertical and constant,
\begin{equation}
 F_0 = -\kappa \frac{\partial T}{\partial z} = {\rm const.} \label{heatflux}
 \end{equation}
The equation of state is that for an ideal gas
and the heat conductivity is constant in the upper and lower layer, respectively, but its values differ between the two layers.

In the dimensionless units  the size of the simulation box is $8\times8\times2$ in the $x$, $y$, and $z$ directions, respectively. The numerical resolution is $512\times512\times128$ grid points.
The stratification of density, pressure, and temperature is piecewise polytrophic as described in Ziegler (2002). Similar setups have been used by Cattaneo et al.~(1991), Brummell et al. (1996), Brandenburg et al.~(1996), Chan (2001) and Ossendrijver et al.~(2001).  
The initial state is in hydrostatic equilibrium but convectively unstable in the upper half of the box. The $z$ coordinate is negative in our setup, with $z=0$ at the upper boundary. The stable layer thus extends from $z=-2$ to $z=-1$, the unstable layer from $z=-1$ to $z=0$. The density varies by a factor 5 over the depth of the box, i.e.~the density scale height is 1.2. 

\begin{figure}[t!]
\begin{center}
\includegraphics[width=\columnwidth]{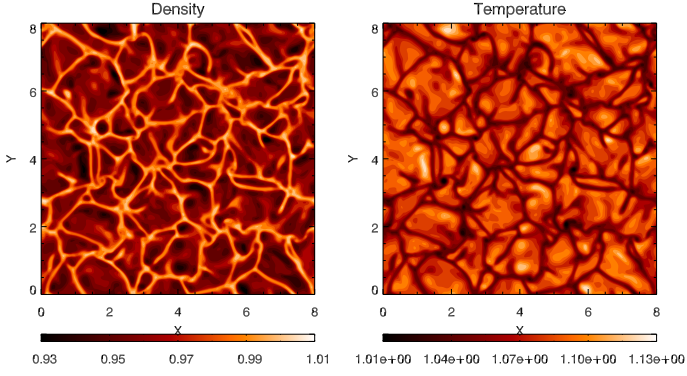}
\end{center}
\caption{
 Fluctuations of density and temperature in the upper part of the  unstable layer at $z=-0.05$ for Ra=$10^7$ and $B_0=10^{-3}$.}
\label{fig1}
\end{figure}

Figure \ref{fig1} shows snapshots of the fluctuations of density and temperature for  ${\rm Ra}=10^7$ in a horizontal plane close to the upper boundary. The density is increased at the boundaries of the convection cells and decreased at the center. The opposite is true for the temperature, which is highest at the center of a convection cell and lowest at the boundaries. Vertical velocity is positive, i.e.~upwards, at the center and negative, i.e.~downwards, at the boundaries. The magnetic field is strongly concentrated in a few small patches which coincide with cell corners, where the gas horizontal flow converges and the vertical flow is downwards.

The initial magnetic field is vertical and homogeneous. We run the simulations until a quasistationary state evolves. Our control parameters are the heat conduction coefficient, $\kappa$ and  the Prandtl number, Pr$=\nu/\kappa$. Convection sets in if the Rayleigh number,
\begin{equation}
  {\rm Ra} = \frac{\rho g c_P d^4}{T \kappa \nu} \left( \frac{dT}{dz} - \frac{g}{c_P} \right),
\end{equation}
with the density $\rho$, the specific heat capacity c$_P$, the gravity force $g$, and the length scale $d$, exceeds a critical value. The length scale is defined by the depth of the convectively unstable layer, i.e.~$d=1$. 
After (\ref{pred})  the   correlations and the mean magnetic field always have opposite signs.
This has  also numerically been  realized. For positive values of the mean magnetic field $ B_z$ the cross helicity is negative in the unstably stratified layer. If the field polarity is reversed and everything else is left unchanged the cross correlation becomes positive with the same amplitude.

\begin{figure}[t!]\begin{center}
\includegraphics[width=\columnwidth]{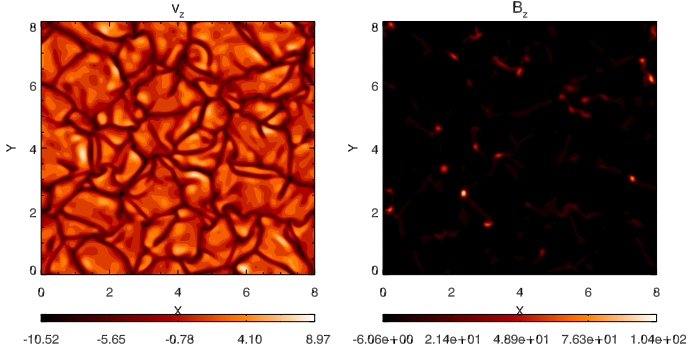}
\end{center}\caption[]{The same as in Fig. \ref{fig1} but for the fluctuations of the vertical flow and the vertical field. $B_0=1$.}
\label{fig2}
\end{figure}

The velocity field, which is measured in units of $c_{\rm ac}/100$, shows the asymmetry between upwards and downwards motion as it is characteristic of convection in stratified media. The downward motion is concentrated at the boundaries of the convection cells and particularly at the corners. The upwards motion fills the interior of the convection cells (see Fig.  \ref{fig2}). As it covers a much larger area the gas motion is much slower than in the concentrated downdrafts. The magnetic field shows a similar pattern. The vertical field is concentrated in the areas  with downwards motion and weak in the areas with upward motion. As the total vertical magnetic flux is conserved,
 this is the result of field advection. 

\begin{figure}[t!]
\begin{center}
\hbox{
\includegraphics[width=4.5cm]{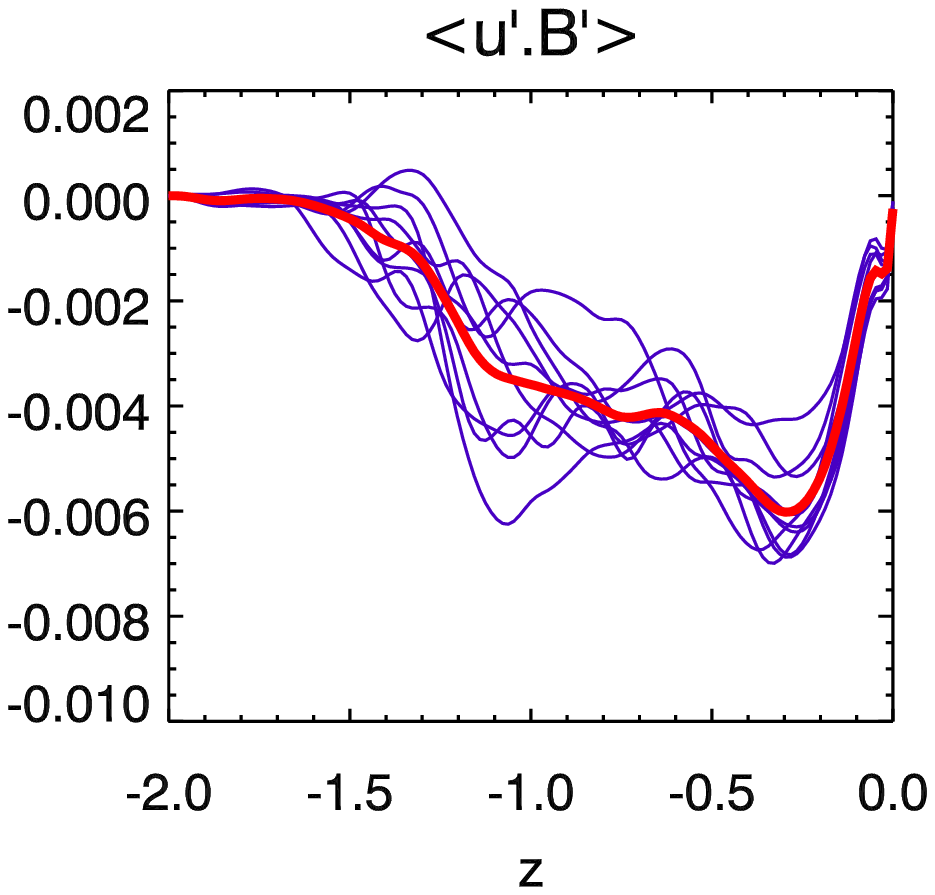}
\includegraphics[width=4.5cm]{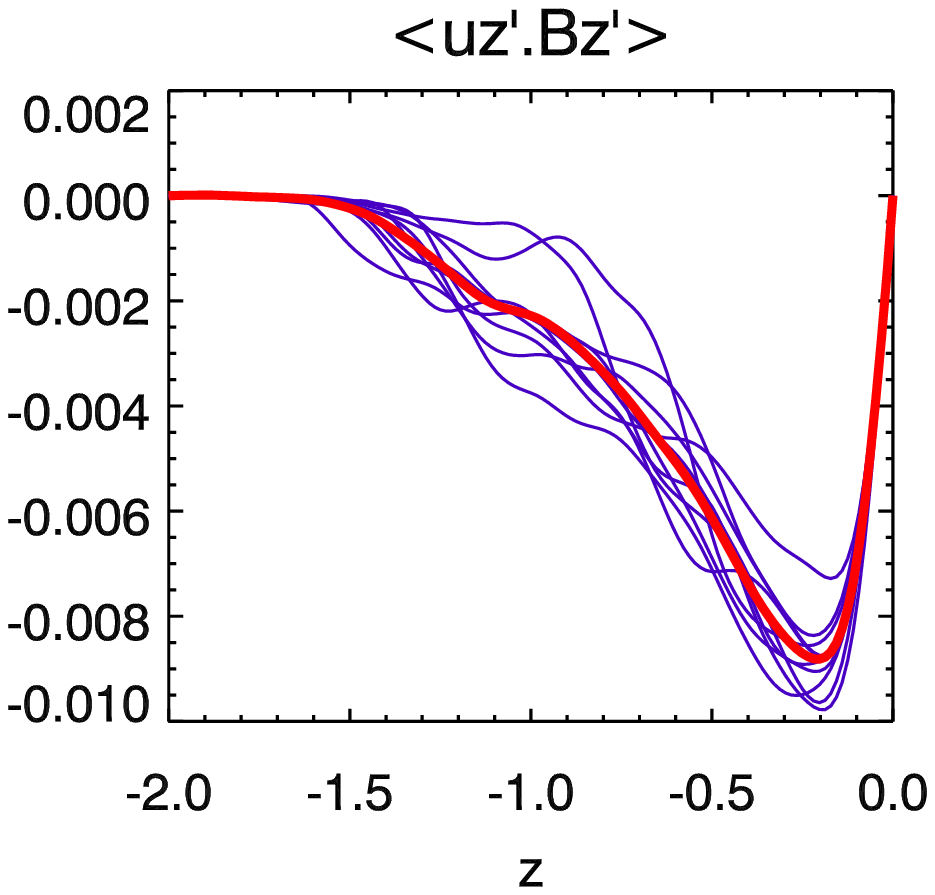}
}
\end{center}\caption[]{
The numerical values for the cross helicity $\langle\vec{ u}\cdot \vec{ b}\rangle$ (left) and the coefficient $\langle u_z b_z \rangle$\ (right) for weak magnetic field, $B_z=10^{-3}$. The blue lines denote individual snapshots and the red lines averages over the snapshots shown. }
\label{fig3}
\end{figure}
\begin{figure}[t!]\begin{center}
\hbox{
\includegraphics[width=4.5cm]{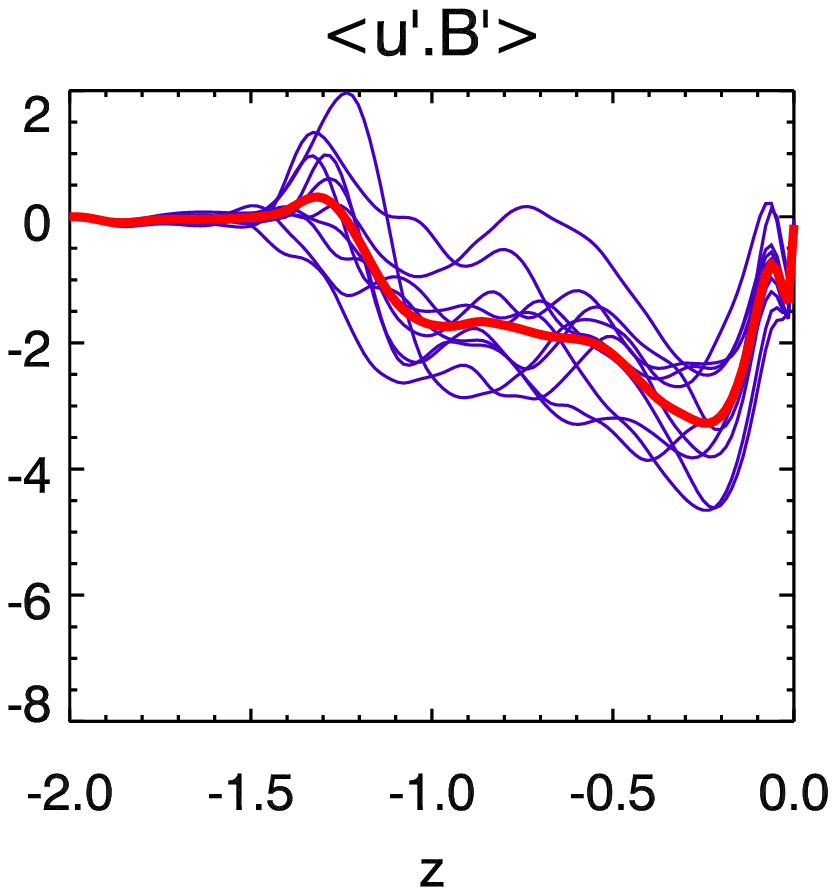}
\includegraphics[width=4.5cm]{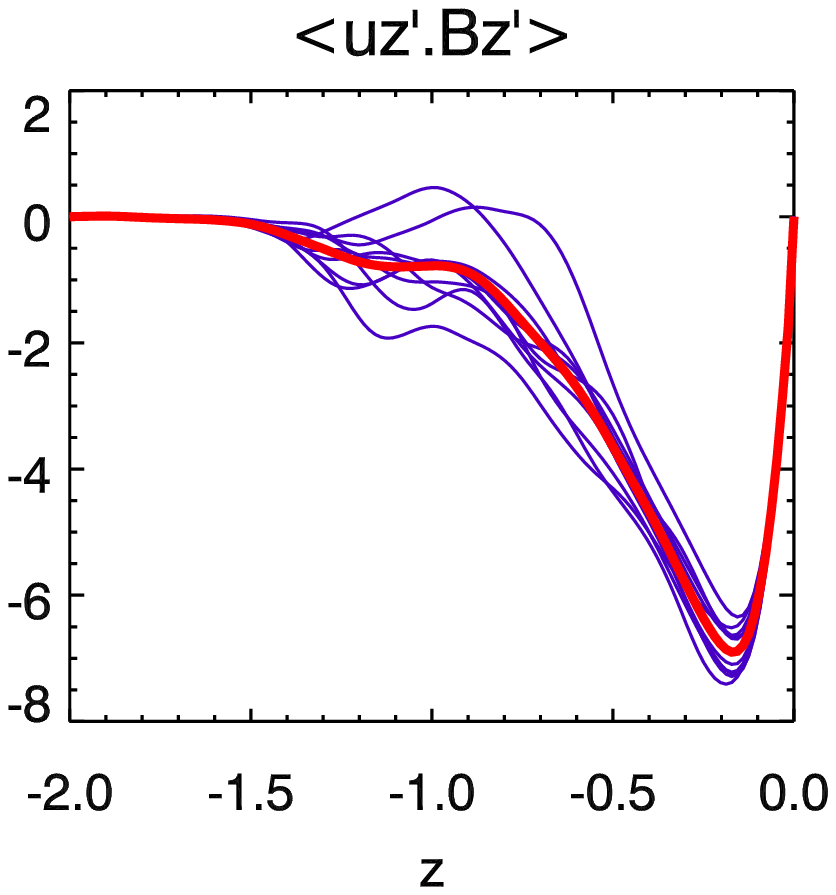}
}
\end{center}\caption[]{The same as in Fig. \ref{fig3} but for  $B_z=1$.
}
\label{fig5}
\end{figure}

\begin{figure}[t!]\begin{center}
\hbox{
\includegraphics[width=4.5cm]{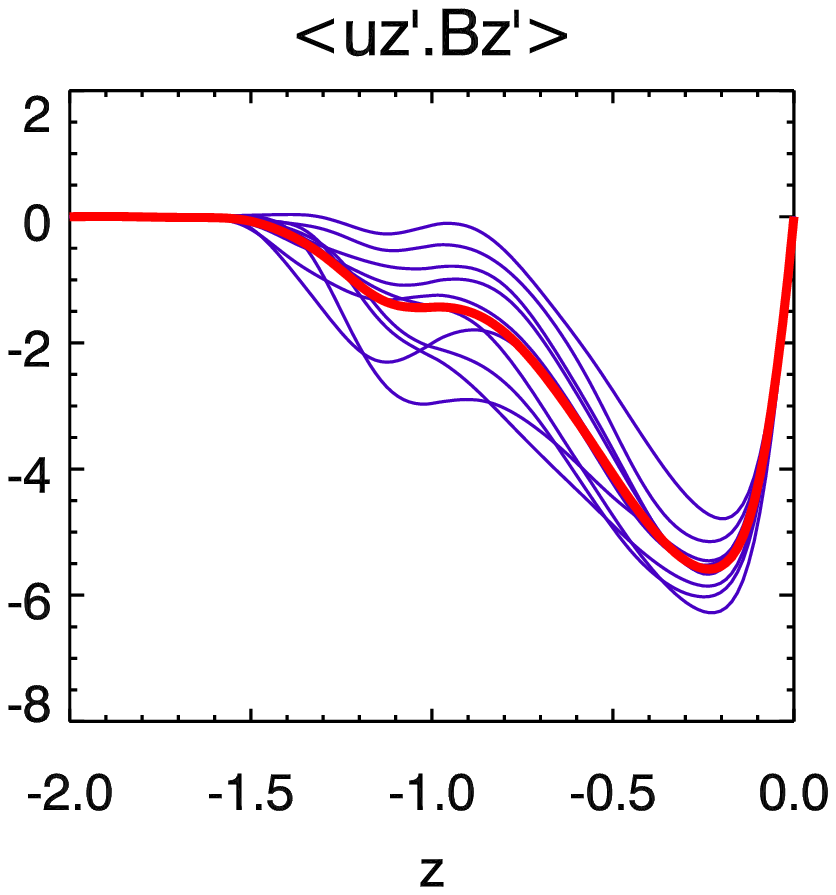}
\includegraphics[width=4.5cm]{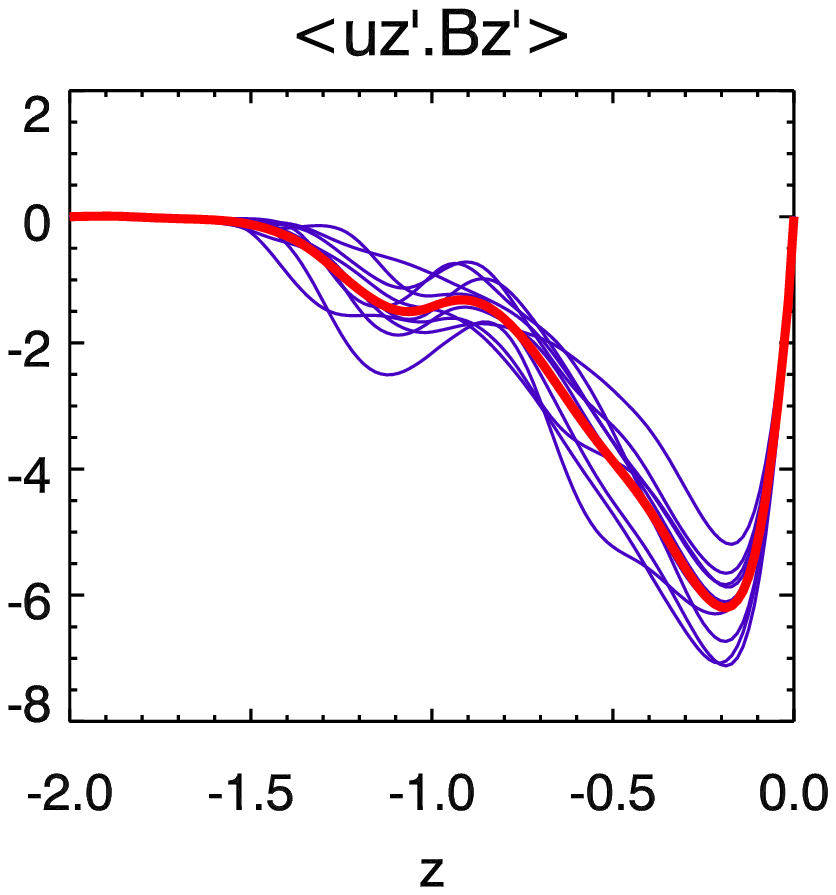}
}
\end{center}
\caption{The same as in the right part of Fig. \ref{fig5}  but with reduced numerical resolution of ($128\times128\times 128$ (left) and $256\times256\times128$ (right).
\label{fig5b} }
\end{figure}

Figures \ref{fig3} and \ref{fig5} hold for  ${\rm Ra}=10^7$ and for  weak and strong magnetic fields. The value of both the Prandtl number and the magnetic Prandtl number is 0.1. The left diagram shows the horizontal average of the cross helicity 
as a function of the depth and the right diagram shows the same for the correlation of the vertical  components, $\langle u_z b_z \rangle$. There is a  difference between the two quantities, with the vertical component actually being twice the  cross helicity. Equation (\ref{12}) can thus be written as 
\begin{equation}
\frac{\langle u_z b_z\rangle}{B_z}\simeq -\frac{4}{7}\ u_\eta  
\label{12a}
\end{equation}
 with
\begin{equation}
u_\eta= \frac{\eta_{\rm T}}{H_\rho} . 
\label{12b}
\end{equation}
The correlations do not vanish abruptly at the bottom of the unstable layer because of overshoot, which affects the upper half of the stable layer. The correlations are there positive and much smaller than in the unstable layer.

The results in Fig. \ref{fig3}  are given in arbitrary units defined by the code. Velocities are given in units of $c_{\rm ac}/100$ with the isothermal speed of sound $c_{\rm ac}$. With an approximate value of $c_{\rm ac}\simeq 6.6$ km/s  at the optical depth $\tau=1$  of the Sun the simulations lead to the cross correlation velocity $\langle u_z b_z\rangle/B_z\simeq - 9$ in units of 0.066 km/s (Fig. \ref{fig3}, right), i.e. after (\ref{12a}) 
\begin{equation}
u_\eta \simeq 1.04\ {\rm km/s} . 
\label{12c}
\end{equation}
This value depends only weakly on the magnetic field amplitude for weak fields. For the much stronger magnetic field Fig.~\ref{fig5} (right) yields the slightly smaller value of 0.81 km/s.

A characteristic velocity results as the cross correlation velocity  
\begin{equation}
U_{\rm c}=\frac{|\langle\vec{u}\cdot \vec{b}\rangle|}{B_z}. 
\label{Ueta}
\end{equation}
 Using the {\em maximal} values in Fig. \ref{fig3} (left) we find $U_{\rm c}\simeq 6$ in units of $c_{\rm ac}/100$. Hence, the simulations lead to the cross correlation velocity $U_{\rm c}\simeq 0.4$ km/s. For the $B_z$=1 case (Fig.\ \ref{fig5}, left) we find $U_{\rm c}\simeq 3$ in units of $c_{\rm ac}/100$ or 0.2 km/s, respectively.
 
It also makes sense to normalize the cross correlation in the form 
\begin{equation}
c_\eta= -\frac{\langle\vec{ u}\cdot \vec{ b}\rangle}{B_z \sqrt{\langle u^2 \rangle }}, 
\label{ceta}
\end{equation}
which is the ratio of the cross correlation velocity (\ref{Ueta}) and the rms velocity of the turbulence. Its numerical value  does not depend on the internal units of the code so that  $c_\eta$ is a general and basic  result of the simulations. Close to the surface the maximal  numerical value  is  $c_\eta\simeq 0.6$ for $B_z=10^{-3}$ and $c_\eta\simeq 0.3$ for $B_z=1$. Test calculations for various magnetic fields over many orders of magnitudes show this value as almost uninfluenced by the magnetic-field suppression. Resulting from the overshoot phenomenon at the bottom of the unstable layer always small  negative values there appear.
The correlation coefficient
\begin{equation}
c= \frac{|\langle\vec{ u}\cdot \vec{ b}\rangle|}{ \sqrt{\langle u^2 \rangle} \sqrt{\langle b^2 \rangle}}, 
\label{c}
\end{equation}
for the cross helicity is much smaller than (\ref{ceta}) as always
\begin{equation}
\frac{{\langle b^2 \rangle}}{B_z^2}\simeq 50, 
\label{cc}
\end{equation}
similar to the result of Ossendrijver et al.\ (2001). The relation (\ref{c}) proves to be true for all amplitudes of the mean magnetic field between $10^{-5}$ and 0.1.
One finds for all calculations a characteristic correlation coefficient $c\simeq 0.1$.
The $B_z=1$ case shows the beginning of the suppression of the fluctuations by the mean magnetic field which occurs at large values of $B_z$, resulting in a smaller value of 25 for
${{\langle b^2 \rangle}}/{B_z^2}$.
\begin{figure}[t!]\begin{center}
\vbox{
\hbox{
\includegraphics[width=4.5cm]{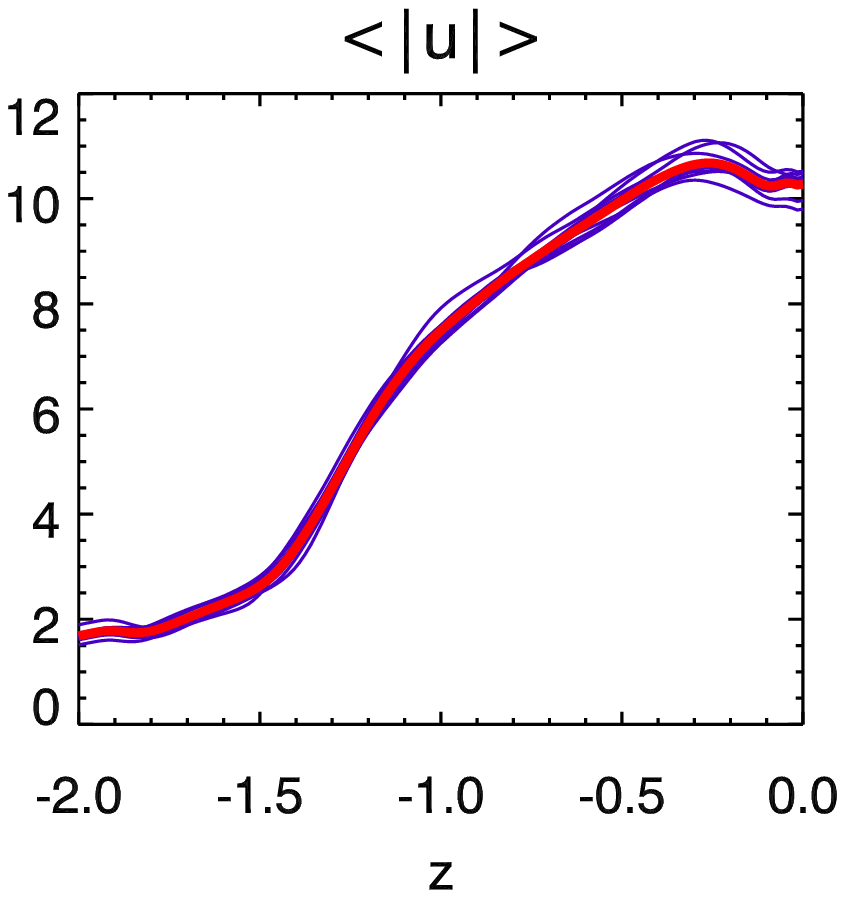}
\includegraphics[width=4.5cm]{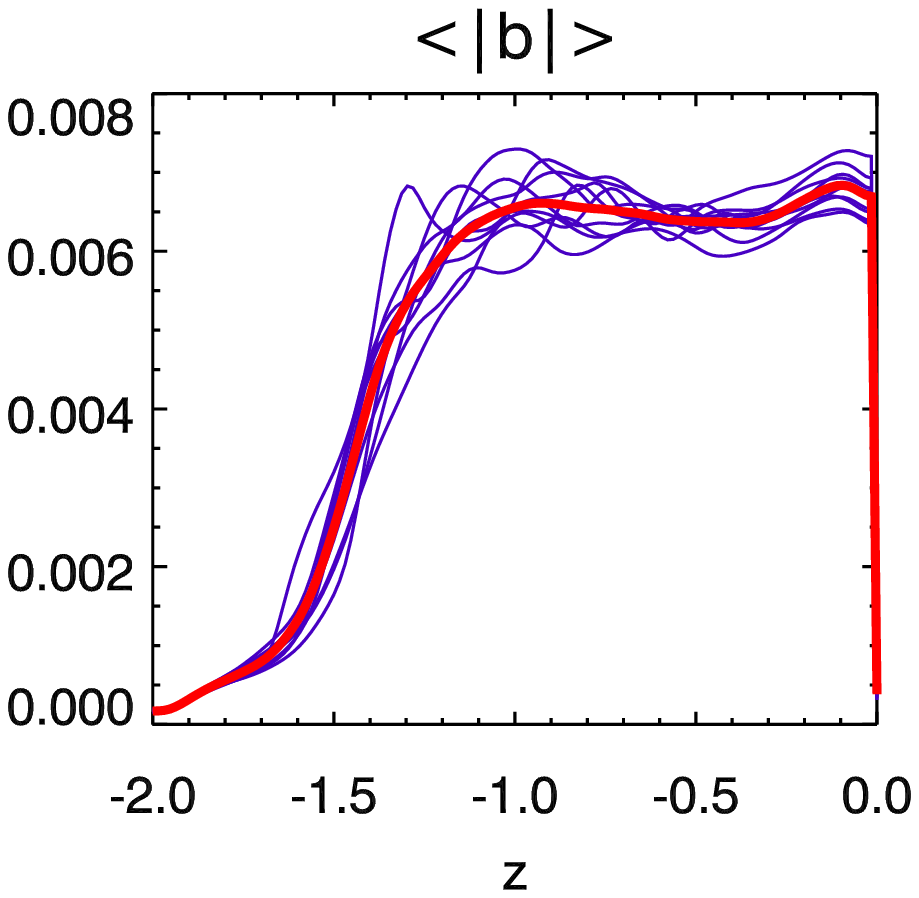}}
\hbox{
\includegraphics[width=4.5cm]{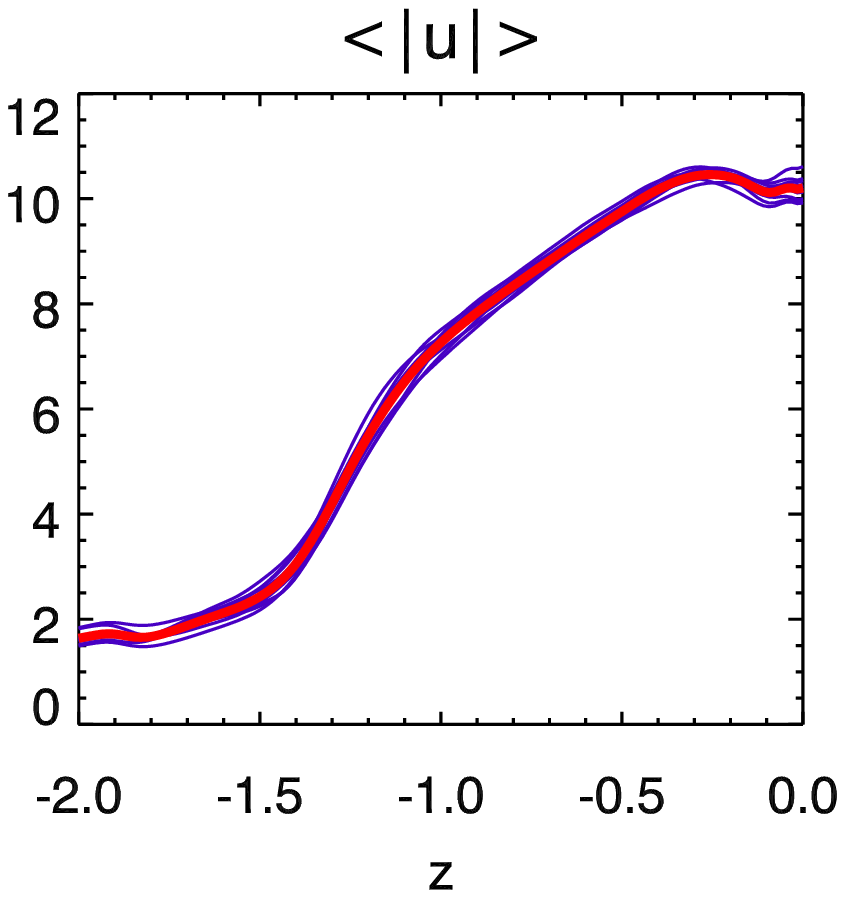}
\includegraphics[width=4.5cm]{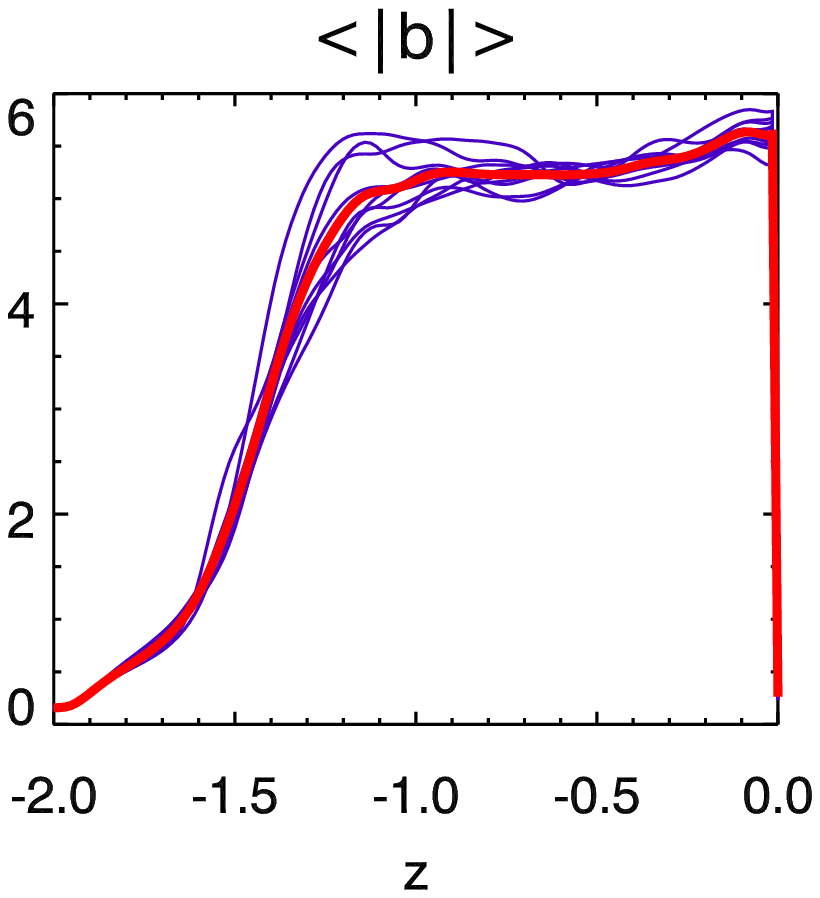}}
}
\end{center}
\caption[]{The numerical values for $u_{\rm rms}$ (left) and for $b_{\rm rms}$ (right) for weak magnetic field, $B_z=10^{-3}$ (top) and strong magnetic field, $B_z=1$ (bottom). Only the magnetic fluctuations depend on the background field amplitude.}
\label{fig4}
\end{figure}

{
During the simulations there are significant temporal fluctuations. The convective instability initially grows exponentially until its saturation when  the system settles in a  statistically steady state but the cross helicity still shows some variations. We therefore average over a certain number of snapshots, typically 10.  
To test how much the results dependend on the numerical resolution we rerun the $B_z=1$ case at the lower lower resolutions of $256\times256\times128$ and $128^3$. Fig.~\ref{fig5b} shows  $\langle u_z b_z \rangle$ from those runs. There is a weak dependence on resolution with higher resolution leading to larger values.
}

Figure \ref{fig4}  contains all informations about the kinetic and magnetic energies of the magnetoconvection. The rms value of the velocity is hardly influenced by the large-scale magnetic field. In physical units we find an averaged value of 
$u_{\rm rms}\simeq 0.1 c_{\rm ac}\simeq 0.66$ km/s. Contrary, the magnetic energy strongly depends on the applied magnetic field. In dimensionless units  it is in both cases $b_{\rm rms}/u_{\rm rms}\simeq 0.6 B_z$ which leads to
\begin{equation}
\frac{E_{\rm mag}}{E_{\rm kin}}\simeq 3600 \frac{B_z^2}{\mu_0\rho c_{\rm ac}^2}, 
\label{energ}
\end{equation}
in physical units. At the top of the convection zone we find very small contributions of the magnetic energy for $B_z=1$ Gauss while for 1000 Gauss there is almost equipartition.

\section{Observations}
It is difficult to empirically determine the cross helicity $\langle\vec{ u}\cdot \vec{ b}\rangle$ at the solar surface, because it is hard to retrieve the horizontal flows and magnetic field components from observations. We have, however, the possibility to use the relation $\langle\vec{ u}\cdot \vec{ b}\rangle\approx 0.5 \langle u_z b_z\rangle$, known from the above numerical simulations. The vertical flow speed and magnetic field component can be determined with much better accuracy. Then using Eq.\ (\ref{12a}) we can determine the cross helicity velocity from the observations.

For this purpose we have analyses two datasets containing observations of quiet Sun at disk center, where the line-of-sight coincides with the local vertical. Data from the CRISP instrument on the Swedish 1-m Solar Telescope (SST) cover the 6302.5 \AA\ \ion{Fe}{i} spectral line with 12 equidistant wavelength positions at 48 m\AA\ steps and a continuum point. They have a pixel scale of $0\, \farcs 0592$ and a total field-of-view of about 60\arcsec$\times$60\arcsec. The second dataset is from the spectropolarimeter on the Solar Optical Telescope of HINODE and covers both the 6301.5 and 6302.5 \AA\ \ion{Fe}{i} lines, has a pixel scale of $0\, \farcs 16$ and a total (scanned) field-of-view of 164\arcsec$\times$328\arcsec. 

The line-of-sight velocity and magnetic field data for the HINODE observations were taken from the level 2 data products available online\footnote{http://sot.lmsal.com/data/sot/level2dd}. Magnetic field strengths have been converted to fluxes by taking the filling factor into account. The SST data were inverted using the lilia inversion code (Socas-Navarro 2001). Velocities were calibrated using the convective blueshift determined by de La Cruz Rodr\'{i}guez et al.~(2011). More details on these two datasets can be found in Schnerr \& Spruit (2011).

We show the results for these datasets in Table \ref{table_observations}. The cross helicity velocity ($u_\eta$) as determined from the SST data is somewhat higher than that from the HINODE data. At least partly this is due to the lower resolution of  HINODE as compared to the SST.  If we rebin the SST data to a lower resolution, the cross helicity velocity decreases (see Table~\ref{table_observations}). The reason for this is that the strongest fields and flows are smoothed out.

{
The coefficient $\langle B_z^2 \rangle / \langle B_z \rangle^2$ from the HINODE and SST data are 521.3 and 163.5 respectively, which is larger than the value of 50 found in the simulations. This indicates that the effective magnetic Reynolds number in the simulations is smaller than in the solar convection zone.}


\begin{table}
\caption{Results from the analysis of the SST and HINODE data. The resolution is measured on the solar surface; the SST resolution is reduced by rebinning the data.}
\centering
\label{table_observations}
\begin{tabular}{lccccc}
\hline\hline
Dataset & Resolution & $\langle u_z b_z\rangle$ & $B_z$ & $\langle u_z b_z\rangle / B_z$ & $u_\eta$\\
& km & [G km/s] & [G] & [km/s] &  [km/s]  \\
\hline
HINODE & 230 & -1.04 & 2.55 & -0.41 & 0.71\\
SST    & 115 & -1.82 & 2.54 & -0.72 & 1.26\\
SST$^*$    & 172 & -1.57 & 2.54 & -0.62 & 1.08\\
SST$^*$    & 258 & -1.32 & 2.54 & -0.52 & 0.91\\
SST$^*$    & 343 & -1.05 & 2.54 & -0.41 & 0.72\\
\hline
\multicolumn{6}{l}{{\footnotesize $^*$These data have been rebinned.}}\\
\end{tabular}
\end{table}



\section{Conclusions}
We  have shown that nonrotating turbulence at the top of the solar
convection zone under the influence of a vertical magnetic field forms
a finite cross helicity.
The only condition is the existence of a vertical  stratification of density and/or turbulence intensity.
The effect would not appear within the Boussinesq approximation.
It also exists in the high-conductivity limit, i.e.\ for
sufficiently large magnetic Reynolds numbers.

In our understanding the  cross helicity  is anticorrelated to the mean radial magnetic field,  i.e.
\begin{equation}
\langle\vec{ u}\cdot \vec{ b}\rangle \cdot    B_z  < 0. 
\label{16}
\end{equation}
For an oscillating dipolar background field the sign of the cross helicity  differs for both hemispheres and also from cycle to cycle.

The theory can also be used to measure the magnetic diffusivity if the cross helicity is known by observations. In order to find the cross helicity one has only to correlate observed flow fluctuations with observed magnetic fluctuations. 

The anticorrelation (\ref{16}) for density-stratified turbulence has been established by R\"udiger et al. (2011) for a model of numerically-driven turbulence. In the present paper buoyancy-driven magnetoconvection has been simulated in a box with the NIRVANA code. We find that also  such a turbulence  fulfills the relation (\ref{16}).  
The correlation coefficient (\ref{ceta}) takes the value of 0.6 for the weak magnetic field $B_z$=$10^{-3}$ and 0.3 for the stronger field $B_z$=1. The ratio (\ref{cc}) of the magnetic fluctuations to the applied magnetic field is always of the order five.

We have also shown that for density-stratified turbulence the identity $\langle\vec{ u}\cdot \vec{ b}\rangle= \langle u_z b_z\rangle$ holds. So far solar observations can only measure the correlation
$ \langle u_z b_z\rangle$. The numerical simulations, however, always lead to the result 
$\langle u_z b_z\rangle \simeq 2 \langle\vec{ u}\cdot \vec{ b}\rangle$ so that the observed value of $ \langle u_z b_z\rangle$ would overestimate the actual cross helicity by a factor of two. The reason is the vertical stratification of the turbulence intensity which at the top of the convection zone is antiparallel to the density stratification. Hence, both the correlations $\langle u_z b_z\rangle$ and $\langle\vec{ u}\cdot \vec{ b}\rangle$ are reduced but not by the same amount. 

With  $\langle u_z b_z\rangle \simeq 2 \langle\vec{ u}\cdot \vec{ b}\rangle$ the value of $u_\eta$ can be computed by use of   Eq. (\ref{12a}). The numerical simulations lead to $u_\eta\simeq 1$ km/s and $u_\eta\simeq 0.8$ km/s respectively for the two cases studied. This result is well confirmed by the observations which lead to values {between} 0.7 km/s (HINODE) and { 1.3} km/s (SST).
To estimate the value of the eddy diffusivity at the solar surface we shall
assume a density scale height of 100 km and find values close to $\eta_{\rm T} \simeq 10^{12}$ cm$^2$/s for the eddy diffusivity at the surface of the {\em quiet} Sun.

%
\acknowledgements
We gratefully acknowledge Axel Brandenburg (Stockholm) for  motivating discussions and for numerical support.

%
%
%

\end{document}